\documentstyle[preprint,aps]{revtex}

\begin{document}
\bibliographystyle{prsty}
\draft

\title{Continued fraction solution to the Raman interaction of
a trapped ultracold ion with two traveling wave lasers}
\author{Mang Feng} 
\address{$^{1}$Max-Planck Institute for the Physics of Complex Systems,\\
N$\ddot{o}$thnitzer Street 38,01187 Dresden, Germany\\
$^{2}$Laboratory of Magnetic Resonance and Atomic and Molecular Physics,\\ 
Wuhan Institute of Physics and Mathematics, Academia Sinica,\\
Wuhan  430071 People's Republic of China}

\date{\today}
\maketitle

\begin{abstract}

Raman interaction of a trapped ultracold ion with two traveling wave lasers
has been used extensively in the ion trap experiments. We solve this
interaction in the absence of the rotating wave approximation by a continued 
fraction, without considering the restriction of the Lamb-Dicke limit and 
the weak 
excitation regime. Some interesting characteristics of the ion-trap system, 
particularly for the ion outside the weak excitation regime are found. 
Finally, a comparison of our results with 
the solution under the rotating wave approximation is made.
\end{abstract}
\vskip 1cm

\narrowtext

Quantum effects in the motion of trapped ultracold ions have been drawn much 
attention 
in recent years since the success that a single ultracold ion was confined
in the ground state of a Paul trap$^{[1]}$. Considerable progress has been made
so far, such as the preparation and measurement of various nonclassical
states$^{[2]}$, and two qubits logic gate operation$^{[3]}$, which are of 
great current interest in ion trap experiment.

It has been proven that a trapped ultracold ion experiencing laser waves can
be described by the Jaynes-Cummings model(JCM)$^{[4]}$ under the
rotating wave approximation(RWA). The JCM, describing a two-level system 
coupled to a harmonic oscillator under RWA in the case of
near resonance and weak coupling, has successfully dealt with the
cavity QED problems, in which a two-level atom interacts with
a monochromatic cavity mode with the form of harmonic oscillator.  In
the trapped ion configuration, the ion is simplified to be two levels
and supposed to move within the region of the space much smaller than the
effective wavelength of the laser wave(Lamb-Dicke limit), the vibrational
level of the trap is quantized as the harmonic oscillator, and the laser
radiating the ion is generally supposed as the classical form of a
standing or traveling wave. Different from the situation of cavity QED, 
the strength of the coupling between the ion and the oscillator can be 
conveniently adjusted simply by changing the intensity of the laser. For
example, in the strong excitation regime$^{[5,6,8]}$, the coupling constant,
denoted by Rabi frequency, is larger than the trapping frequency, which
makes RWA no longer valid. 

Under RWA, along with the assumption that the ion is confined within
the Lamb-Dicke limit(LDL) and weak excitation regime(i.e., the Rabi
frequency being smaller than the trapping frequency), one may easily
obtain three simple cases$^{[4,6]}$, that is, carrier excitation, red detuning
and blue detuning. If excluding the supposition with respect to the LDL
and weak excitation regime, Vogel and his coworkers gave a standard
approach to such a nonlinear JC model$^{[7]}$. However, in case the RWA is
excluded, the ion trap problem turns to be non-integrable, whose solutions
have to be obtained completely by the numerical calculation$^{[6,8]}$. 
Recently, we 
analytically treated$^{[9]}$ the interaction in the absence of RWA between 
a standing wave laser and a trapped ultracold ion in coherent state 
representation$^{[10,11]}$. Although those analytical results are only some 
particular solutions to the problem, some important information neglected in 
former works due to use of RWA were presented. However, our investigation in 
that work was only under both the weak excitation regime and the LDL. Moreover,
the solutions in that work strongly depended on the truncation of the series, 
and thereby the deduction and the forms of the analytical expressions went 
tedious with the increase of the number of terms in the series, which makes it 
hard to obtain the eigenenergies and eigenfunctions rapidly at our disposal. 
Furthermore, it is Raman configurations that are usually applied in actual ion 
trap experiments for the interaction between lasers and ions, instead
of the simple case described in Ref.[9]. Therefore, in this contribution, we 
choose Raman $\Lambda$-type configuration(shown in Fig.1) for a study,
which corresponds to the actual process in the NIST experiments$^{[1]}$. To
present a more general description for such a model, we will exclude 
the assumption related to both the weak excitation regime and LDL by
means of some unitary transformations. Moreover,
to overcome the difficulty of complicated analytical deduction, our solution 
will resort to a continued fraction, from which the
eigenenergies of the system and the coefficients of the series of the
eigenfunctions can be rapidly obtained by numerical calculation.

Fig.1 describes the interaction of a trapped ultracold ion with two
off-resonant counter-propagating traveling wave lasers with frequencies
$\omega_{1}$ and $\omega_{2}$ respectively. Both laser beams are assumed to 
propagate along the $x$ 
axis, and the problem is thereby one-dimensional. For a sufficiently
large detuning $\delta$, the third level $|r>$ may be adiabatically eliminated,
and we face an effective two-level system, in which the lasers drive the
electric-dipole forbidden transition $|g>$$\leftrightarrow |e>$. In the 
frame rotating with the effective laser frequency $\omega_{L}$
$(=\omega_{1}-\omega_{2})$, the dimensionless Hamiltonian of this
system is$^{[1,2,12]}$
\begin{equation}
H=\frac {\Delta}{2}\sigma_{z} + a^{+}a+ \frac {\Omega}{2}
(\sigma_{+}e^{i\eta \hat{x}}+\sigma_{-}e^{-i\eta \hat{x}})
\end{equation}
where $\Delta=(\omega_{0}-\omega_{L})/\nu$ with $\omega_{0}$ being the resonant
frequency of the two levels of the ion, and $\nu$ being
the trap frequency. $\Omega$ is the dimensionless Rabi frequency and $\eta$
the Lamb-Dicke parameter. $\sigma_{i}$ ($i=\pm, z$) are Pauli operators,
$\hat{x}=a^{+}+a$ is the dimensionless position operator of the ion with
$a^{+}$ and $a$ being operators of creation and annihilation of the phonon 
field, respectively. The notations '+' and '-' in front of $i\eta \hat{x}$ 
indicate that the laser wave propagates towards $x$ and -$x$ directions, 
respectively. $\nu$ is generally supposed to be much greater 
than the atomic decay rate(strong confinement limit) for neglecting the effect 
of the atomic decay.

We first perform some unitary transformations on Eq.(1)$^{[12]}$, and obtain
\begin{equation}
H^{I}=UHU^{+}=\frac {\Omega}{2}\sigma_{z} + a^{+}a+ g(a^{+}+a)(\sigma_{+}+
\sigma_{-})+\epsilon (\sigma_{+}+\sigma_{-})+g^{2}
\end{equation}
where $U=\frac {1}{\sqrt{2}}e^{i\pi a^{+}a/2}\pmatrix{D^{+}(\beta)
& D(\beta) \cr -D^{+}(\beta)& D(\beta)}$ with $D(\beta)=e^{i\eta(a^{+}+a)/2}$,
$g=\eta/2$ and $\epsilon=-\Delta/2$. 
Comparing with Ref.[9], Eq.(2) can be regarded mathematically as the 
simplified form of Eq.(6) in Ref.[9] without the two-phonon processes,
whereas the optical resonance frequency is replaced by the Rabi frequency. 
Similar to Ref.[9], we set $a^{+}\rightarrow \alpha$, $a\rightarrow
\frac {d}{d\alpha}$, and rewrite Eq.(2) in coherent state representation, 
that is,  
\begin{equation}
H={\frac {\Omega}{2}}\sigma_{z}+ \alpha {\frac {d}{d \alpha}}+
g(\alpha+\frac {d}{d \alpha})(\sigma_{+}+\sigma_{-})+
\epsilon (\sigma_{+}+\sigma_{-})+g^{2}
\end{equation}
where $\alpha$ is a complex number and $\int \frac {d\alpha d\alpha^{*}}
{2\pi i} \exp (-|\alpha|^{2}) |\alpha^{*}><\alpha^{*}|=1$. The eigenfunction 
of the Schr\"odinger equation of Eq.(3) can be supposed to be 
$\Psi(\alpha)= \pmatrix{\Psi_{1}(\alpha) \cr\Psi_{2}(\alpha)}$, and the
eigenenergy is E. Thus we have
\begin{equation}
(\alpha+g)\frac {d}{d\alpha}\Phi_{1}=(E-g^{2}-g\alpha-\epsilon)\Phi_{1}-\frac
{\Omega}{2}\Phi_{2},
\end{equation}
\begin{equation}
(\alpha-g)\frac {d}{d\alpha}\Phi_{2}=(E-g^{2}+g\alpha+\epsilon)\Phi_{2}-\frac
{\Omega}{2}\Phi_{1}
\end{equation}
where we have made a transformation $\Phi_{1}=\Psi_{1}(\alpha)+
\Psi_{2}(\alpha)$ and $\Phi_{2}=\Psi_{1}(\alpha)-\Psi_{2}(\alpha)$. To solve 
Eqs.(4) and (5), we order $\xi=\alpha+g$,
and $\Phi_{1}=\exp{(-g\xi)}\phi (\xi)$. Combining Eq.(4) with Eq.(5) will
yield a second-order differential equation 
$$\xi (\xi -2g)\frac {d^{2}}{d\xi^{2}}\phi +[2gE-2g-2g\epsilon+(1+4g^{2}
-2E)\xi -2g\xi^{2}]\frac {d}{d\xi}\phi + $$ 
\begin{equation}
[E^{2}-\epsilon^{2}-\frac {\Omega^{2}}{4}-4g^{2}E+4g^{2}\epsilon
+2g(E-\epsilon)\xi]\phi=0.
\end{equation}
By supposing $\phi$ to be a series form $\phi=\sum^{\infty}
_{n=0} C_{n}\xi^{n}$, Eq.(6) becomes a recurrence relation about
the coefficient $C_{n}$
\begin{equation}
\alpha_{n}C_{n+1}+\beta_{n}C_{n}+\gamma_{n}C_{n-1}=0
\end{equation}
where $\alpha_{n}=2g(n+1)(E-\epsilon-1-n)$, $\beta_{n}=[n^{2}+2n(2g^{2}-E)
+E^{2}-\epsilon^{2}-\frac {\Omega^{2}}{4}-4g^{2}E+4g^{2}\epsilon]$ 
and $\gamma_{n}=2g(E-\epsilon-n+1)$. Obviously, by cutting the series
at $C_{-1}=0$, we have $C_{1}/C_{0}=-\beta_{0}/\alpha_{0}$, and other
coefficients $C_{n} (n=2,3,\cdots)$ will be denoted by $C_{0}$. Making
a slight transformation for Eq.(7), we obtain following continued fraction,
\begin{equation}
\beta_{n}-\frac {\gamma_{n}\alpha_{n-1}}{\beta_{n-1}-
\frac {\gamma_{n-1}\alpha_{n-2}}{\beta_{n-2}-\cdots\frac {\gamma_{2}\alpha_{1}}
{\beta_{1}-\gamma_{1}\alpha_{0}/\beta_{0}}}}
=\frac {\alpha_{n}\gamma_{n+1}}{\beta_{n+1}-\frac {\alpha_{n+1}\gamma_{n+2}}
{\beta_{n+2}-\frac {\alpha_{n+2}\gamma_{n+3}}{\cdots}}}
\end{equation}  
from which the eigenenergies E can be easily obtained, and the
coefficients $C_{n}(n=1,2,\cdots)$ can also be solved$^{[13]}$.

Prior to the numerical calculation for Eq.(8), we first consider a
special case of the problem, that is, the case of LDL and large
detuning in the strong excitation regime. At this time, Eq.(3)
reduces to the mathematical form similar to the carrier excitation, i.e., 
$H={\frac {\Omega}{2}}\sigma_{z}+ \alpha {\frac {d}{d \alpha}}+
\epsilon (\sigma_{+}+\sigma_{-})$. With above procedure, we can obtain
$$\beta_{n}=0\Longrightarrow E=n\pm\sqrt{\frac {\Omega^{2}}{4}+\epsilon^{2}}$$
which is in good agreement with the solution in Fock state representation.
Except for this special case, the other situations of the problem have
to be solved numerically from Eq.(8). Obviously, the convergence of the
coefficients $C_{n}$ is the necessary condition in the calculation. As this
condition has been discussed in Ref.[13], we will not repeat it here. 
The numerical calculation of Eq.(8) demonstrated in Figs.2,3 and 4 may
present us following interesting results:\\
(i) the eigenenergies of the system increase with the enhancement of $\Omega$
and $\eta$, which means that it is more difficult to stably confine the
ultracold ion outside the LDL and weak excitation regime; \\
(ii) the increase of the eigenenergies with the enhancement of $\eta$ is
suppressed by the large detunings. In view of physics, the only
reasonable explanation for this case is that the interaction between
the laser and the ion plays main role in the system. With the increase
of the detuning, this interaction goes weaker and weaker. Therefore,
although the energy of the ion itself is enhanced with the increase of $\eta$,
the energy of the total system remains nearly constant;\\
(iii) Fig.4 presents the relation between $\Omega$ and $\Delta$ for certain 
values of E and $\eta$.
From the forms of the curves, we may guess that the relation among E, $\Delta$,
$\eta$ and $\Omega$ is of the form of $\Omega=f(E)[1-h(\eta)\Delta^{2}]$ with 
$f(E)$ and $h(\eta)$ being positive definition functions of E and $\eta$, 
respectively.\\
Moreover, from the specific calculation, we found that, for a certain $\Omega$,
with the increase of $\eta$, the convergence of $C_{n}$ would get worse
and worse, particularly for the large detuning case. But the larger the
value of $\Omega$, the larger the value of $\eta$ for satisfying the
convergence of $C_{n}$. The problem with respect to the convergence
will be discussed physically later. 

Returning Eqs.(4) and (5), we can find that $\Phi_{1}=\exp{(g\xi)}\phi$ 
is another solution for $\xi=\alpha +g$. Repeating above deduction, we
will obtain Eq.(8), whereas in this case, $\alpha_{n}=2g(n+1)(E-1-\epsilon
-n)$, $\beta_{n}=(n^{2}-2nE-4ng^{2}+E^{2}-\epsilon^{2}-\frac {\Omega^{2}}{4}
-4g^{2})$ and $\gamma_{n} = 2g(n-\epsilon-E)$. Similar to
above procedure, we studied the relation among E, $\Delta$, $\Omega$
and $\eta$, as shown in Figs.5, 6 and 7. Comparing with Figs.2, 3 and 4 
respectively, we may find that the values in Figs.5, 6 and 7 are slightly 
different, and their difference become more and more obvious with the increase 
of $\eta$ and $\Omega$, or E and $\eta$. Particularly for the case of 
$\Omega=2$ and $\Delta=2.0$ in Fig. 6, the curve no longer upgrades with the
increase of $\eta$.

As we obtained the solutions of $\Phi_{1}$, $\Phi_{2}$ can be easily
solved from Eqs.(4) and (5) with the form of $\Phi_{2}=-\frac {2}{\Omega}
\sum_{n=0}^{\infty}(n-E+\epsilon)C_{n}(\alpha +g)^{n}e^{\pm g(\alpha+g)}$.
Therefore, the eigenfunction of the system is
\begin{equation}
\Psi(\alpha)= \pmatrix{\Psi_{1}(\alpha) \cr\Psi_{2}(\alpha)}=
\frac {1}{2} e^{\pm g(\alpha +g)}
\pmatrix{\sum_{n=0}^{\infty}C_{n}[1-\frac {2}{\Omega}(n-E+\epsilon)]
(\alpha +g)^{n} \cr \sum_{n=0}^{\infty}C_{n}[1+\frac {2}{\Omega}(n-E+\epsilon
)](\alpha +g)^{n}}
\end{equation}
From Refs.[9,10,15], we know that the coherent states can be conveniently
transformed to the Fock states with the formula
\begin{equation}
|n>=\int \frac {d\alpha d\alpha^{*}}{2\pi i} 
\exp (-|\alpha|^{2}) |\alpha^{*}>\frac {1}{\sqrt{n!}} \alpha^{n}.
\end{equation} 
By means of Baker-Campbell-Hausdorff formula, we may find
$e^{g\alpha}\alpha^{n}$ in coherent state representation correspond to the
displaced Fock state $|n,g>$. Therefore, $\Psi (\alpha)$ in Eq.(9) is in fact
a superposition of several displaced Fock states. However, to obtain the
eigenfunctions in the original representation, we should
consider the transformation made in Eq.(2). As an example, we may treat
the special case $H={\frac {\Omega}{2}}\sigma_{z}+ 
\alpha {\frac {d}{d \alpha}}+ 
\epsilon (\sigma_{+}+\sigma_{-})$, whose eigenfunction is 
\begin{equation}
\Psi^{S}(\alpha)= \pmatrix{\Psi^{s}_{1}(\alpha) \cr\Psi^{s}_{2}(\alpha)}=
\frac {1}{2} \pmatrix{A \cr B} \sum_{n=0}^{\infty}C_{n}\alpha^{n}
\end{equation}
where $A=1\pm \frac {2}{\Omega}\sqrt{\frac {\Omega^{2}}{4}+\epsilon^{2}}+
\frac {2\epsilon}{\Omega}$, $B=1\mp \frac {2}{\Omega}\sqrt{\frac {\Omega^{2}}
{4}+\epsilon^{2}}+ \frac {2\epsilon}{\Omega}$. Then the eigenfunction
in the original representation is 
\begin{equation}
\begin{array}{l}
\Psi^{SO}(\alpha)= \frac {1}{\sqrt{2}}e^{i\pi a^{+}a/2}
\pmatrix{1&1 \cr -1&1}\frac {1}{2}\pmatrix{A\cr B}|n>\\
=\frac {1}{\sqrt{2}}e^{in\pi/2}\pmatrix{1+\frac {2\epsilon}{\Omega} \cr 
\pm\frac {2}{\Omega}\sqrt{\frac {\Omega^{2}}{4}+\epsilon^{2}}}|n>
\end{array}
\end{equation}
where we have made $\alpha^{n}$ to be Fock state $|n>$ by means of Eq.(10), and
simply chosen the coefficients related to $\alpha^{n}$ to be 1 and 
$\alpha^{m}$$(m\neq n)$ to be zero. From above
equation we know that, in the 
weak excitation regime $\Omega\ll\epsilon$,
or in the carrier excitation, i.e.,
$\epsilon =0$, $\Psi^{SO}(\alpha) \propto \pmatrix{1 \cr \pm 1}|n>$, which is 
identical to the standard solution from Fock state representation$^{[6]}$.
In general, in the strong excitation regime with large detuning, the 
probability amplitudes of up and down states will be different, as shown in Eq.(12). 

As referred to previously, the problem of eigenfunction and eigenenergy
for a trapped and radiated ion
was treated numerically in Ref.[6], in which the Hamiltonian was diagonalized 
in the space spanned by the dressed Fock states. As the solutions of the 
non-RWA JC model are with the forms of coherent states or squeezed states, 
etc$^{[9,14,15]}$, the accurate eigenenergies and eigenfunctions in Ref.[6] 
might not be in principle solved unless the space for diagonalization was 
spanned by the infinite dressed Fock states, particularly in the case of the 
large detuning and strong excitation regime. That is the reason why Ref.[6] 
only presented the solutions in the weak excitation regime 
for the case of near-resonance. In our treatment, however,  
the numerical solution is made based on the continued fraction, which is
obtained in the coherent state representation. So our solution is
still valid even for the large detuning and strong excitation regime. 

Before ending our discussion, it is interesting to make a comparison
between our solution with those under RWA. 
In general, we may solve Eq.(1) by means of nonlinear JC model as
proposed in [7] for the ion governed outside the weak excitation
regime. However, to make the comparison easily and obviously, we start
our comparison procedure from Eq.(2). By performing a unitary
transformation $\exp [-i(\frac {\Omega}{2}\sigma_{z}+\frac {a^{+}a}{2})t]$
on Eq.(2), we can obtain $E^{\pm}_{n}=\frac {2n+1}{4}+\frac {1}{4}\eta^{2}
\pm\frac {1}{4}\sqrt{4\eta^{2}(n+1)+1}$ $(n=0, 1, 2, \cdots)$ in the Fock 
state representation for 
$\Omega=2$ under RWA(see Figs.8). From Figs.3, 6 and 8, we see that, 
for small values of $\eta$, the case of $n=2$ under RWA is similar to those 
of $\Omega=2$ in 
Figs.3 and 6, where $E_{n}^{+}$ and $E_{n}^{-}$ are similar respectively to 
the very large and zero detuning cases without RWA. Although both the
solutions with RWA and without RWA include multi-curve, the difference
is obvious: as it is merely a particular solution, the solution without
RWA approaches to the solution of a certain excited state under RWA. On the
other hand, as RWA merely retains some resonance terms in Hamiltonian, 
the solution under RWA is irrelative to the variation of $\Delta$ and
just related to a specific value of $\Omega$. Moreover, 
as they are obtained from the solution of Eq.(8), the results without RWA are 
restricted by the convergence of the coefficients of the
series. In contrast, there is no restriction mathematically for the
expressions of the solution under RWA even in the case of
$\eta\rightarrow\infty$. However, in physics, $\eta$ is impossible to
approach to infinity for a ultracold ion system, and it must be cut off
at a certain value. Therefore, in this sense, the convergence condition
in our solution of the continued fraction provides a physical restriction on 
the final results, which plays the same role as the restricted condition
for the solutions in Ref.[9]. Furthermore, further deduction with the
transformations
$\exp [-i(\frac {\Omega}{2}\sigma_{z}+\frac {a^{+}a}{4})t]$ and $\exp
[-i(\frac {\Omega}{2}\sigma_{z}+\frac {a^{+}a}{6})t]$ performed on Eq.(2) 
respectively 
can tell us that, with the increase of $\Omega$ and $\eta$, the difference
between the solutions with RWA and without RWA will be larger and larger.

In summary, we have presented a continued fraction solution to a trapped and 
laser-radiated ultracold ion experiencing two traveling wave lasers in
a Raman process in the absence of RWA. Instead of the complicated 
and tedious analytical forms in Ref.[9], we can solve the problem 
rapidly from the continued fraction and present more complete information of
the system, such as the relation among E,
$\Omega$, $\Delta$ and $\eta$. As limitation of weak excitation
regime and LDL as well as RWA is excluded in our treatment, our 
solution is not only exact, but more general. It can be used to investigate
quantum properties of the ion-trap system in a wider range of parameters,
and serve as a comparison with other approximate works, as we did in Ref.[16].
However, as Eq.(2) represents a kind of typical non-integrable system, our 
solutions are still only some particular ones. While in comparison with 
former
solutions under RWA, we still found some interesting results, particularly
 for 
the case outside the weak excitation regime and LDL.
We should emphasize that, it is hard to carry out the trapped 
ion-laser interaction experimentally outside the weak excitation regime at
present due to the requirement of very high 
intensity of the laser, and the laser cooling to the ground state of the
trapping potential has not been reported yet in this regime. Therefore,
we hope the solution in our work would be helpful for the future exploration
of the quantum properties of the ion-trap system in this respect.

The work is partly supported by National Natural Science Foundation of
China under Grant No. 19904013.

$Note~added$: after accomplishing this paper, the author becomes aware of that a work$^{[17]}$
related to the fast quantum gate for cold trapped ions has been carried out based on Ref.[12]
and the treatment similar to our RWA treatment.

\begin{center} {\bf Captions of the figures} \end{center}

Fig.1~~ Level scheme of the internal structure of the trapped ultracold
ion, where $|g>\leftrightarrow |e>$ is dipole forbidden.

Fig.2~~ Variation of E with respect to  $\Delta$ in the case of 
$\Phi_{1}=\exp{(-g\xi)}\phi$, where the dotted, dashed, solid and
dash-dotted curves represent $\eta=$0.2, 0.4, 0.6 and 0.8, respectively.
For clarity of the figure, we only plot the cases for $\Omega=$2, 4 and 6.

Fig.3~~ Variation of E with respect to  $\eta$ in the case of 
$\Phi_{1}=\exp{(-g\xi)}\phi$, where the solid, dotted, dash-dotted
and dashed curves represent $\Delta=$0.2, 1.6, 2.0 and 3.0, respectively.
For clarity of the figure, we only plot the cases for $\Omega=$2, 4 and 6.

Fig.4~~ Variation of $\Omega$ with respect to  $\Delta$ in the case of 
$\Phi_{1}=\exp{(-g\xi)}\phi$, where the solid, dashed and dotted curves 
represent $\eta=$0.02, 0.4 and 0.6, respectively. We only plot the cases for 
$E=$3 and 5.

Fig.5~~ Variation of E with respect to  $\Delta$ in the case of 
$\Phi_{1}=\exp{(g\xi)}\phi$. The rest are the same as that in Fig.2.

Fig.6~~ Variation of E with respect to  $\eta$ in the case of 
$\Phi_{1}=\exp{(g\xi)}\phi$. The rest are the same as that in Fig.3.

Fig.7~~  Variation of $\Omega$ with respect to  $\Delta$ in the case of 
$\Phi_{1}=\exp{(g\xi)}\phi$. The rest are the same as that in Fig.4.

Fig.8~~  Variation of $E^{\pm}_{n}$ with respect to $\eta$ for $\Omega=2$ 
for the solution under RWA. See text.
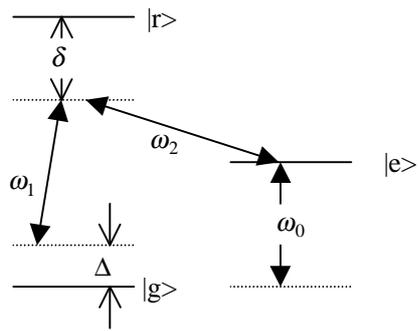
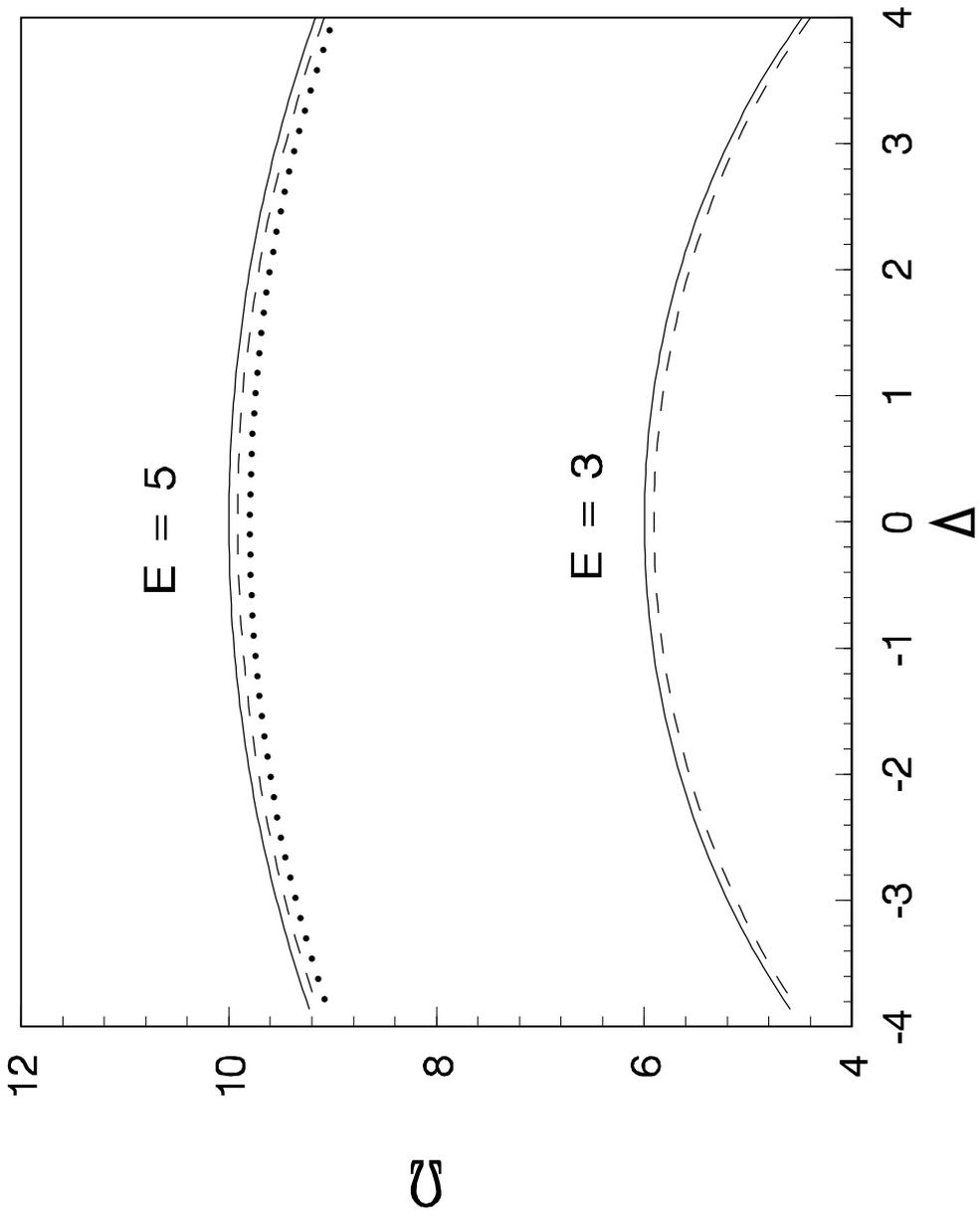
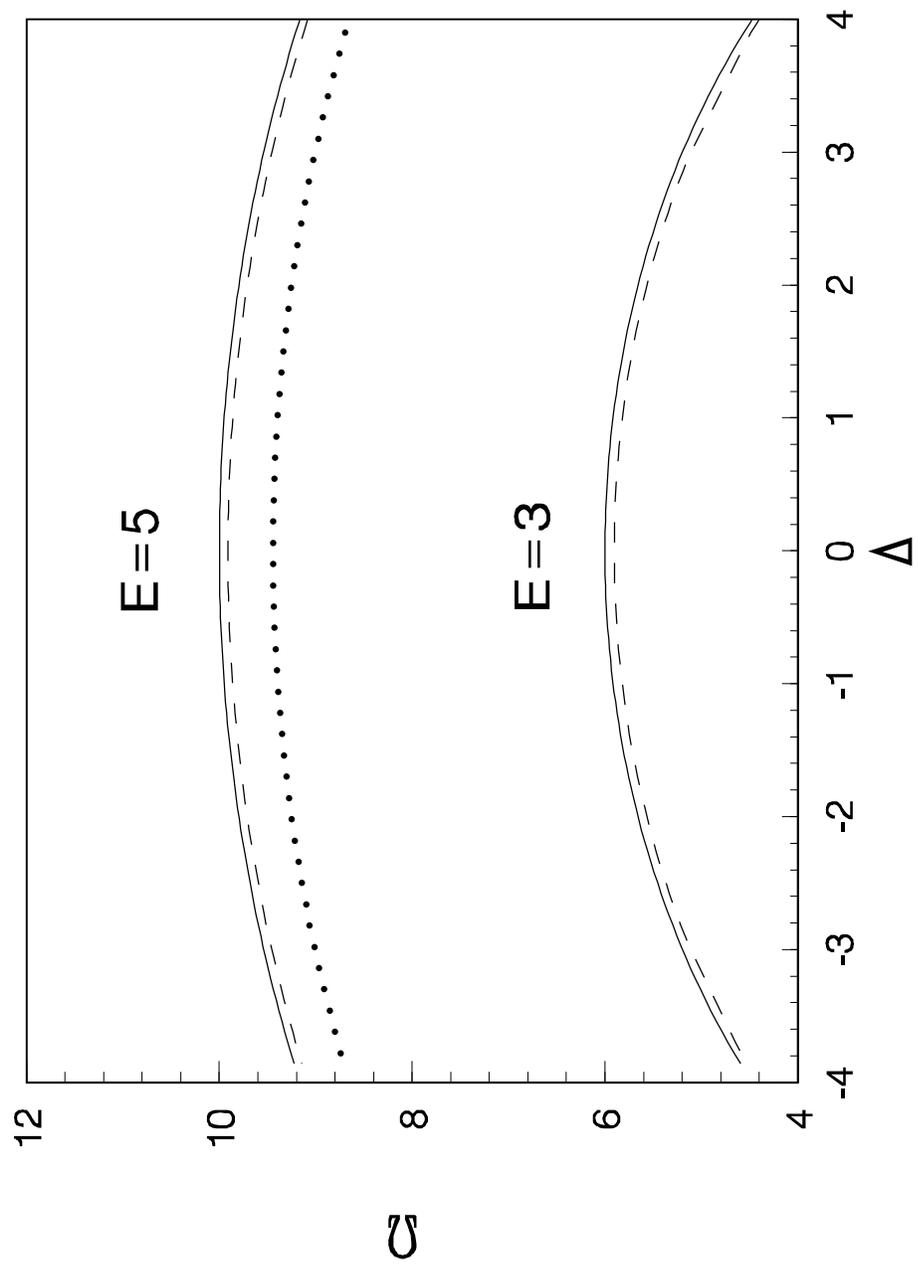
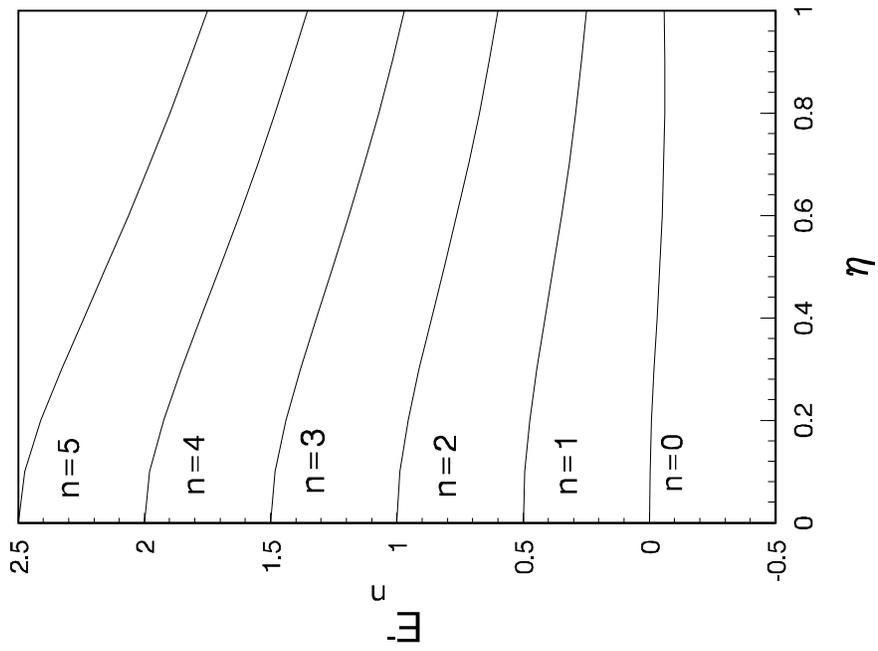
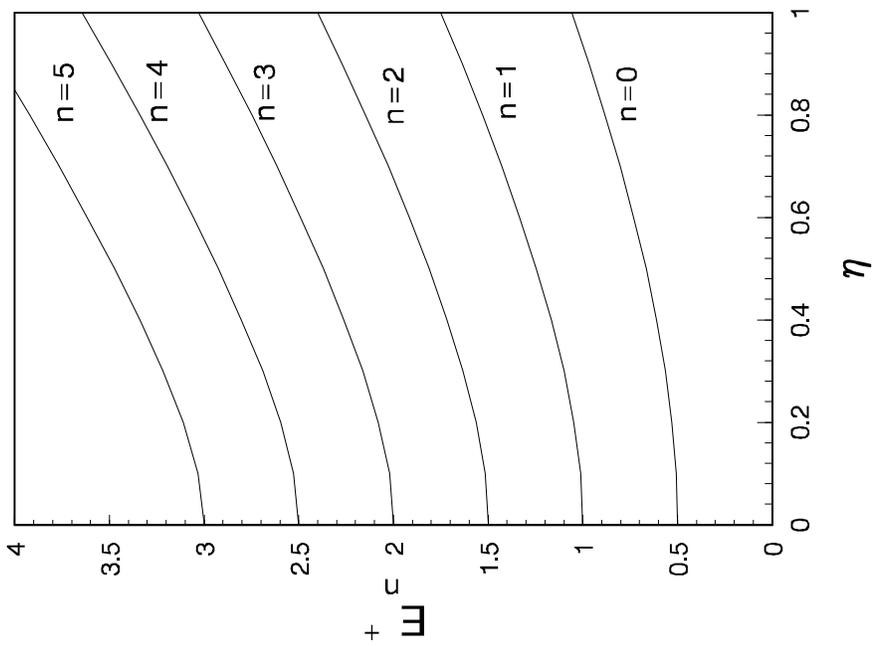
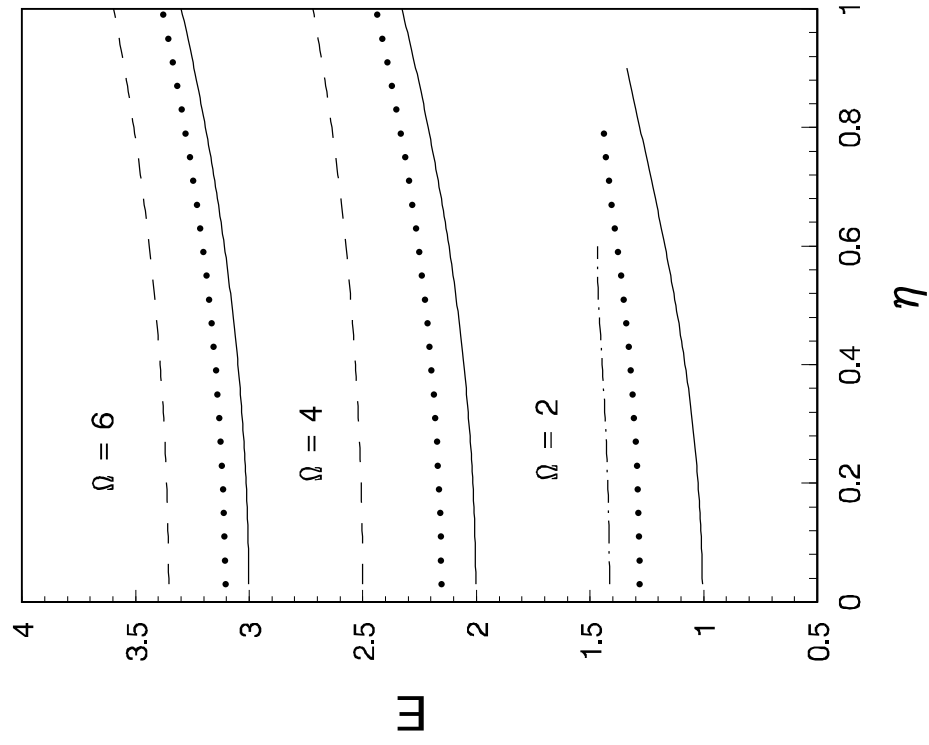
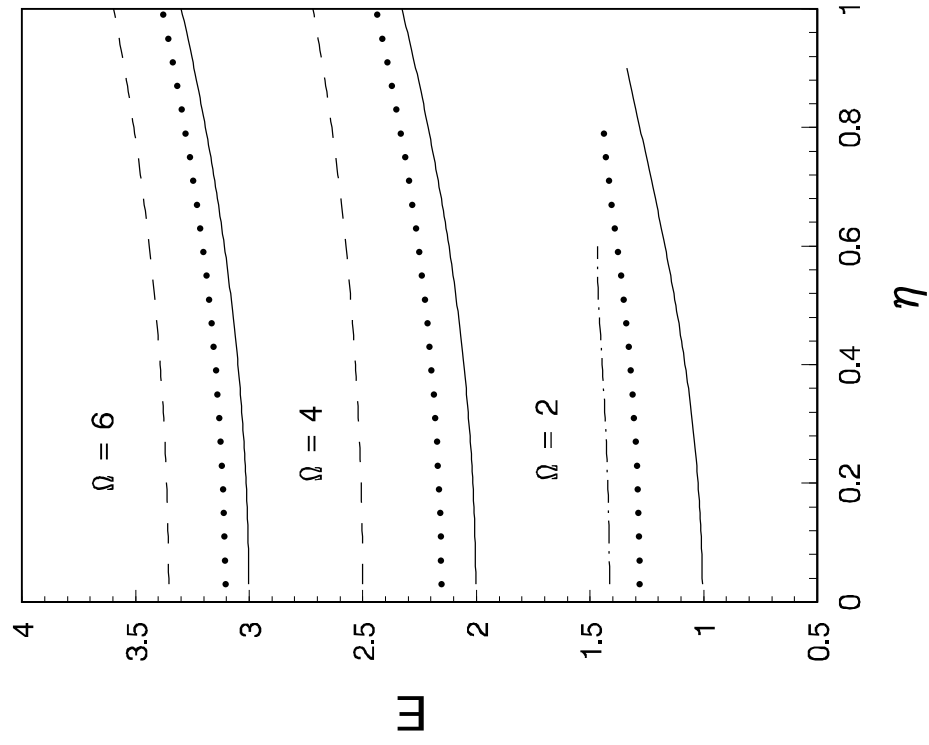
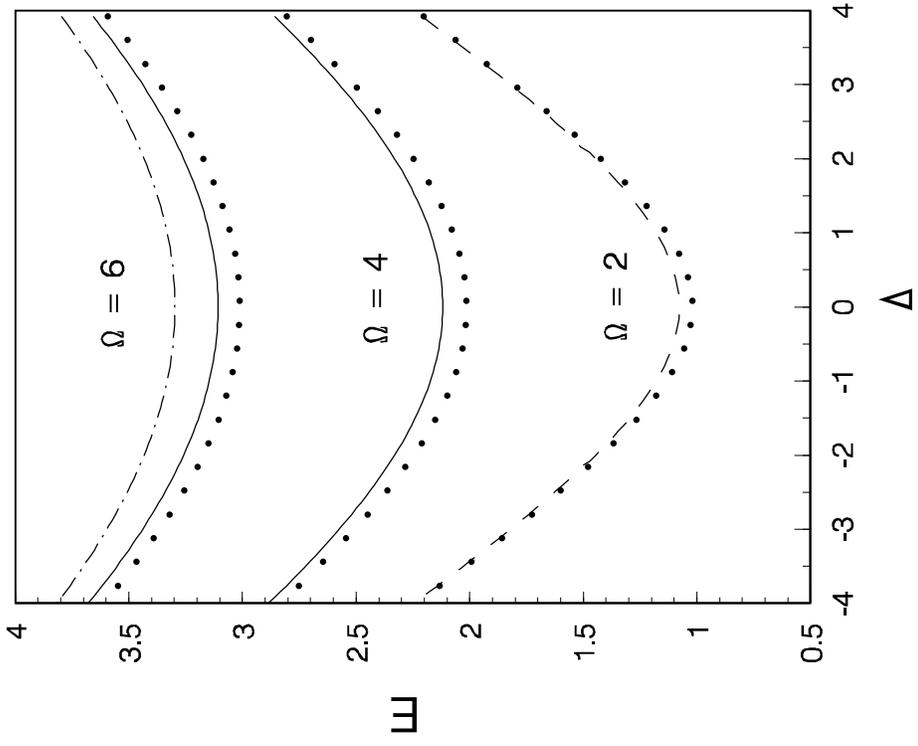
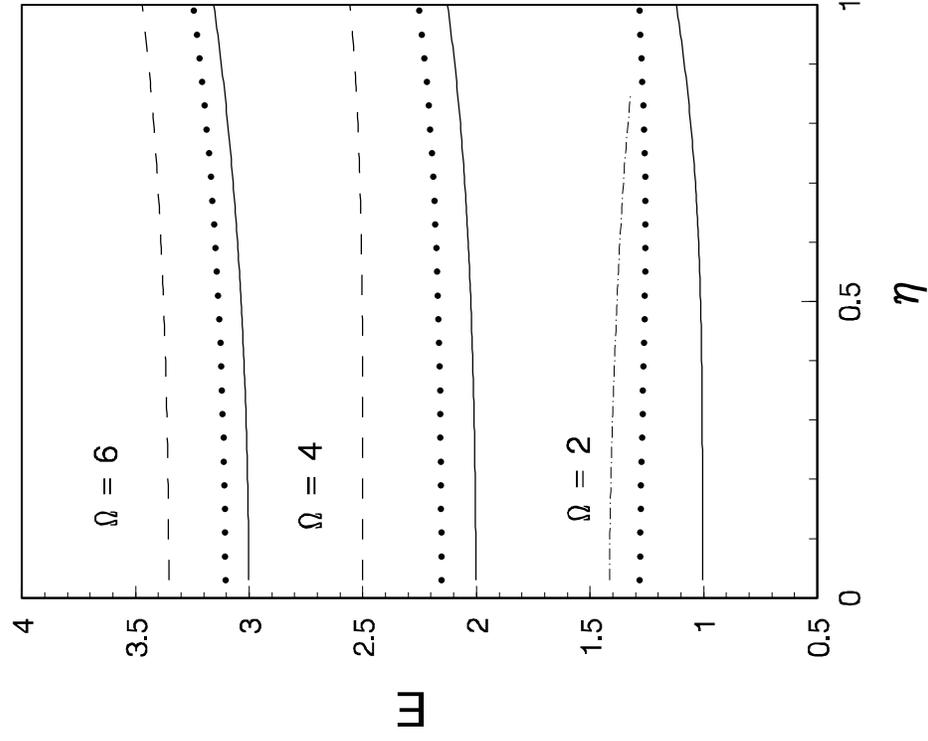
\end{document}